\pdfoutput=1

\documentclass[11pt]{article}

\usepackage[final]{acl}
\usepackage{booktabs}
\usepackage{times}
\usepackage{mathptmx} 
\usepackage{latexsym}
\usepackage{adjustbox}
\usepackage{amsmath}  
\usepackage{amssymb}  
\usepackage{multirow}
\usepackage{array}
\usepackage{graphicx} 
\usepackage{float} 
\usepackage{algorithm}
\usepackage{listings}
\usepackage{algpseudocode}
\usepackage{xspace}
\usepackage{graphicx}
\usepackage{etoolbox}
\usepackage{xcolor}
\usepackage{tcolorbox}
\usepackage{hyperref}
\usepackage{latexsym}

\usepackage[T1]{fontenc}

\usepackage[utf8]{inputenc}

\usepackage{microtype}

\usepackage{inconsolata}

\usepackage{graphicx}

\usepackage{microtype}
\newcommand{\system}{\emph{GraPPI}\xspace}
\newcommand{\IG}{\emph{Interaction Graph}\xspace}
\newcommand{\IS}{\emph{Impact Search}\xspace}

\title{GraPPI: A Retrieve-Divide-Solve GraphRAG Framework for Large-scale Protein-protein Interaction Exploration}

\author{Ziwen Li \\
  UCLA\\
  Los Angeles, USA\\
  \texttt{zil105@ucla.edu} \\\And
  Xiang `Anthony' Chen \\
  UCLA\\
  Los Angeles, USA\\
  \texttt{xac@ucla.edu} \\\And
  Youngseung Jeon\footnote{*} \\
  UCLA\\
  Los Angeles, USA\\
  \texttt{ysj@ucla.edu} \\}

\newcommand\blfootnote[1]{%
  \begingroup
  \renewcommand\thefootnote{}\footnote{#1}%
  \addtocounter{footnote}{-1}%
  \endgroup
}

\begin{document}
\maketitle

\begin{abstract}
Drug discovery (DD) has tremendously contributed to maintaining and improving public health. Hypothesizing that inhibiting protein misfolding can slow disease progression, researchers focus on target identification (Target ID) to find protein structures for drug binding. While Large Language Models (LLMs) and Retrieval-Augmented Generation (RAG) frameworks have accelerated drug discovery, integrating models into cohesive workflows remains challenging. We conducted a user study with drug discovery researchers to identify the applicability of LLMs and RAGs in Target ID. We identified two main findings: 1) an LLM should provide multiple Protein-Protein Interactions (PPIs) based on an initial protein and protein candidates that have a therapeutic impact; 2) the model must provide the PPI and relevant explanations for better understanding. Based on these observations, we identified three limitations in previous approaches for Target ID: 1) semantic ambiguity, 2) lack of explainability, and 3) short retrieval units.
To address these issues, we propose \system, a large-scale knowledge graph (KG)-based retrieve-divide-solve agent pipeline RAG framework to support large-scale PPI signaling pathway exploration in understanding therapeutic impacts by decomposing the analysis of entire PPI pathways into sub-tasks focused on the analysis of PPI edges~\footnote{\textcolor{blue}{Code link:} \url{https://github.com/AaronLi43/GraPPI}}\blfootnote{*Corresponding author.}.

\end{abstract}

\section{Introduction}

\begin{figure}[h]
    \centering
    \includegraphics[width=\linewidth]{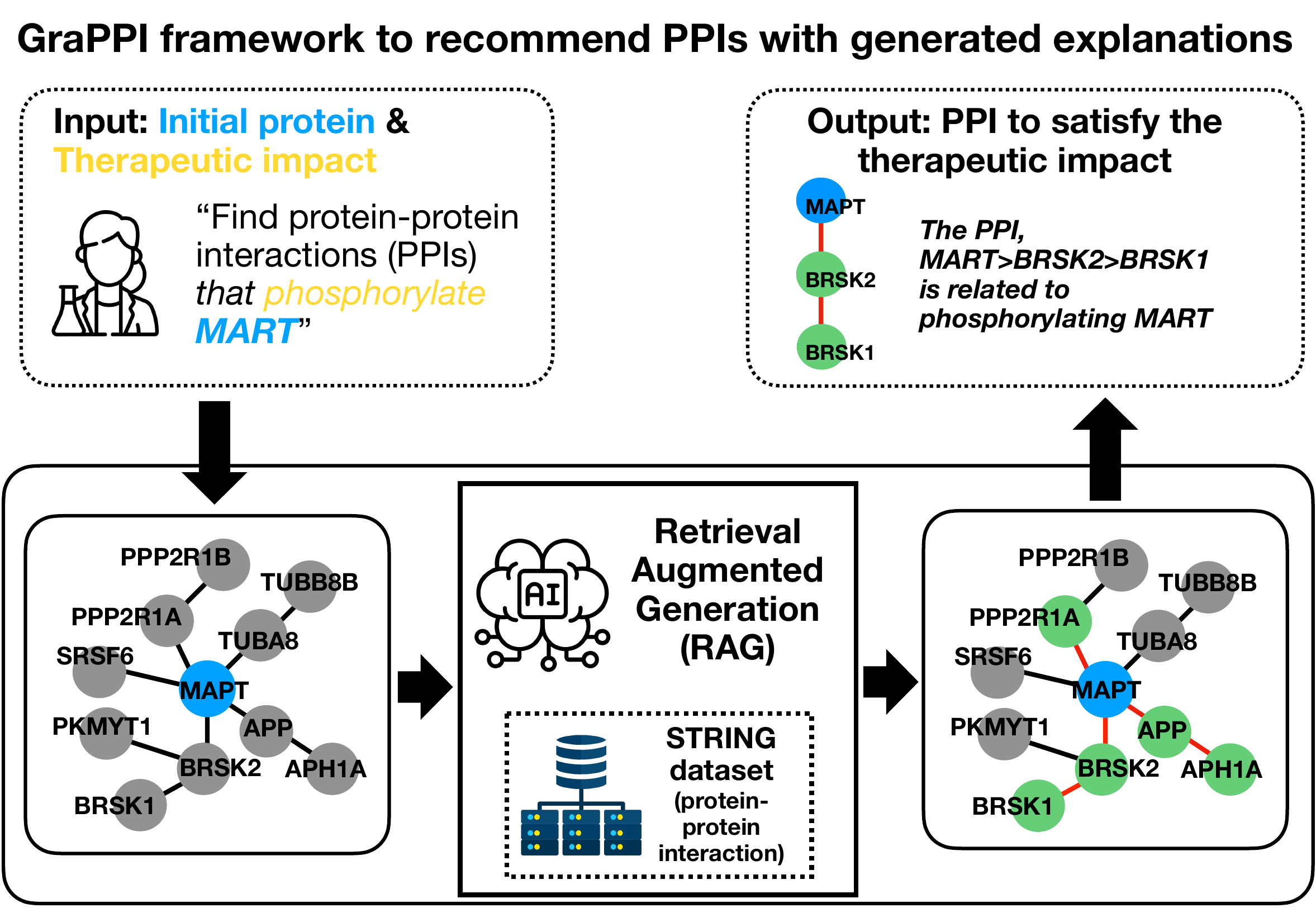}
    \caption{GraPPI for target identification (Target ID). Based on the two inputs- an initial protein and therapeutic impact on the initial protein- GraPPI recommends PPI pathways with explanations and retrieves text based on previous work.}
    \label{fig:demo}
\end{figure} 
The discovery of new drugs can potentially create treatments that save lives and enhance health outcomes globally. For example, penicillin, discovered in the early twentieth century, revolutionized bacterial infection treatments and saved countless lives~\cite{drews2000drug}. Based on the critical hypothesis that inhibiting and activating protein misfolding can slow disease progression, drug discovery (DD) researchers are focused on \textit{Target identification (Target ID)}, the process of elucidating the protein structures that drugs can bind to. As depicted in Figure~\ref{fig:demo}, this process aims to identify protein-protein interactions (PPIs), which are protein pathways from an initial protein (IP) to protein candidates for the target protein (TP). The TP should have a therapeutic impact on the IP. Given that the number of protein candidates in the human body is several billion~\cite{smith2013proteoform}, \textit{Target ID} is very time-consuming and expensive, requiring DD researchers to explore PPI candidates within the extensive protein space by scanning related literature for validation.
\begin{figure*}[h]
    \centering
    \includegraphics[width=\linewidth]{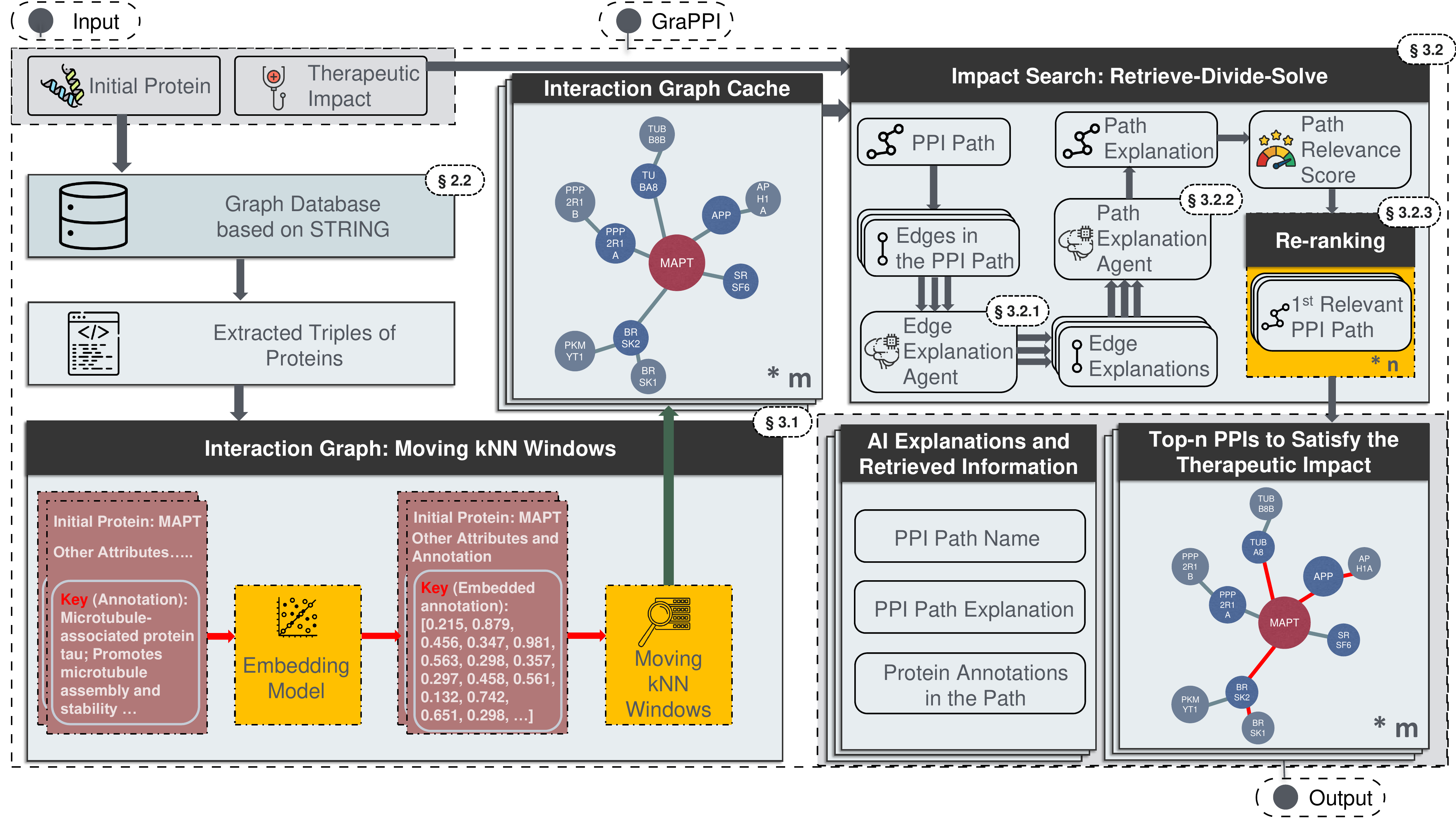}
    \caption{Overview of \system framework. The input of the users to \system is the name of the initial protein and the therapeutic impact query. The outputs are recommended PPIs with AI-generated explanations and retrieved information from the database. }
    \label{fig:overview}
\end{figure*}

Recently, Large Language Models (LLMs) \cite{openai2023chatgpt, touvron2023llama, devlin2019bert, google2024gemini} have become a trending generative model for understanding user intents \cite{li2023gpt4rec, zhao2023recommender, kang2023llms} such as preferred therapeutic impacts and generating biomedical advice in natural language \cite{bender2020climbing, vaswani2017attention}. However, they are prone to hallucinations that introduce unreliable results \cite{ji2023survey, zhang2023siren}. As a result, Retrieval-Augmented Generation (RAG) represents a promising solution by combining a constantly updated database with efficient information retrieval for more accurate and contextually-related responses \cite{lewis2020retrieval, li2023understand, khandelwal2019generalization}. After recognizing the potential to mitigate hallucination and improve the reliability in generative models \cite{martineau2023retrieval, zhang2023knowledge}, RAG has been introduced to the biomedical-related realms \cite{lin2024generag, wang2024biorag, li2024biomedrag, yang2024harnessing} where reliability and explainability matter especially for domain experts.

To identify the applicability of LLM- or RAG-based methods, we conducted user studies with five drug discovery researchers.
We observed that an LLM should provide multiple Protein-Protein Interactions (PPIs) based on initial protein and protein candidates that align with the therapeutic impact. Additionally, the model must provide the PPI and relevant explanations for researchers’ understanding. Based on these observation, we identified three main limitations in applying LLMs to \textit{Target ID}.

\textbf{First}, previous works have semantic ambiguity issues where the subject and object shift \cite{ji2023survey}. RAG may learn or predict incorrect relationships if it fails to interpret this accurately and may struggle to recognize relation information from a corpus of drastically increasing biomedical literature before PPI signaling pathway analysis, due to noise or irrelevant information \cite{ji2023survey}. Knowledge Graphs (KGs) are widely used to store structured information with well-assigned attributes to each entity. This can mitigate inaccurate data interpretation of the used dataset in the RAG database \cite{wang2023boosting}.


\textbf{Second}, the lack of explainability of LLM-based and RAG-based systems when reasoning and inference process hinders fact-checking and exacerbates over-reliance on AI recommendations \cite{si2023large}. RAG typically only provides retrieved information for the generators rather than to users. Domain experts cannot easily access to the related context to verify the truthfulness of their expertise.

\textbf{Third}, 
Short retrieval units in existing RAG frameworks restrict performance of tasks requiring extended context information, particularly in exploring pathways with multiple PPIs, while general KGs usually have shorter contexts \cite{sun2024oda, luo2023reasoning}. A limited context window leads to semantic incompleteness due to truncation of supporting information \cite{jiang2024longrag}. 
To bridge these gaps, we propose \system, with the key idea illustrated in Figure \ref{fig:overview}, a KG-based RAG framework that supports the exploration of PPI signaling pathways. We incorporate all \textit{Homo sapiens} PPI data from the STRING dataset to construct our KG in \system. To guarantee data reliability in the used dataset, we consulted with DD domain experts, who confirmed full confidence in the STRING dataset. It utilizes fast \textit{k}-nearest neighbor (\textit{k}NN) search windows to extract relevant PPI context in the entire KG and breaks down the task of long PPI signaling pathways inference into parallel sub-tasks focused on PPI edges inference in the potential pathways using a retrieve-divide-solve style agent pipeline to \textbf{incorporate long contexts of protein annotations to PPI exploration to support fact-checking.} For LLM inference on single PPIs and entire PPI signaling pathways, we introduce co-designed Chain-of-Thought (CoT) with domain experts to help agents to understand single PPI analysis and multi-PPI pathway analysis. 



In summary, the key contributions of our work are as follows:

\begin{itemize}
    \item We propose \system, a KG-based RAG framework in a retrieve-divide-solve agent style pipeline to break down the exploration of the entire PPI pathway into parallel analysis of sub-edges inside to enable accurate inference on long PPI contexts.
    \item We propose a hybrid retrieval strategy incorporating multiple fast \textit{k}NN searches and LLM inference, which promotes efficient fact-checking based on contexts retrieved in large-scale KG for fact-checking and mitigate over-reliance on AI recommendations.
    \item To our best knowledge, we have constructed the largest KG covering all human PPI information for improved interpretation of entity relationships in a reliable dataset, consisting of 18,767 proteins and 2,955,220 PPIs from the STRING dataset.
\end{itemize}





\section{User Study}
In this appendix, we describe the details of our user study (Section~\ref{userstudy-approach}), existing processes in Target ID (Section~\ref{userstudy-result}), and the challenges and applicability of RAG as potential solutions (Section~\ref{userstudy-applicability}).


\subsection{Approach}\label{userstudy-approach}
We recruited five researchers specializing in drug discovery and conducted user studies in February 2024 to gain a deep understanding of their \textit{Target ID} process. Their work experience spans 7 to 10 years (mean = 8.75, SD = 1.3). Each interview lasted approximately 60-90 minutes. Three authors attended all sessions, took notes during the discussions, and later consolidated and analyzed the notes in wrap-up meetings. Two participants, who were our collaborators, were not compensated, while the remaining three received a \$30 gift certificate after the study. Table~\ref{tab:demo_p} in the appendix shows participants' demographic details.

During the interviews, we asked about (1) their current practices regarding \textit{Target ID} and (2) the applicability of LLMs in \textit{Target ID}. After completing the interviews, we employeed thematic analysis and iterative open coding~\cite{clarke2015thematic} to analyze the interview transcripts. Three researchers coded and analyzed the transcripts for emerging themes, and the findings were iteratively discussed among the co-authors until reaching a consensus.

\subsection{Results}\label{userstudy-result}

\textit{Target identification (Target ID)} is introduced as a process to explore the protein space for PPI signaling pathways. PPI signaling pathways are paths of proteins that start from an initial protein (IP) that is therapeutically related to certain diseases to a target protein (TP). There are two conditions for TP. First, the target protein should have physical and functional interactions ranging from itself to the initial protein (C1). To identify these interactions, scientists input the initial protein on STRING~\footnote{\url{https://sea.bkslab.org/}}, and then they retrieve a protein interaction graph consisting of hundreds of proteins based on the initial protein. Before moving on to the next step, the scientists want to identify as many proteins as possible to increase the likelihood of finding the optimal target protein. Secondly, scientists look for possible therapeutic impacts to make an initial protein be inhibited or activated (C2). Scientists search for the therapeutic impacts of proteins on the interaction graph via Google Search. For example, MAPT is a key protein associated with Alzheimer's disease because excessive phosphorylation (activation) of MAPT promotes the disease. When researchers search for therapeutic impacts of proteins, they should verify whether a protein phosphorylates MAPT. As a result, scientists filter the proteins on the graph to retain only those with the desired impact, significantly reducing their number.


\subsection{Applicability}\label{userstudy-applicability}
The space of possible targets is expansive, given that researchers estimate around 10,000~\cite{adkins2002toward} proteins in the human body. Scientists are constrained to exploring limited protein candidates, negatively impacting scientific discoveries. In our user study, the experts provided highly positive feedback on using LLMs with PPI graphs to explore the extensive protein space, citing their efficiency in identifying potential therapeutic target protein (TP) candidates. If LLMs can identify PPIs having desired TPs in a PPI graph based on an initial protein (IP) consisting of several PPIs, it demonstrates their strong potential to support scientific discovery effectively. However, they also emphasized the necessity of utilizing a confident dataset, such as the STRING dataset, that provides interactions for over 3 million proteins. Researchers also mentioned the importance of fact-checking with explanations for why the model recommends specific PPIs. The absence of scientific materials for results could diminish the quality of LLM-generated recommendations.


In summary, key findings for supporting Target ID are as follows:

\begin{itemize}
    \item  A RAG should provide a PPI composed of multiple proteins, with the IP and TP as start and end points, respectively, and ensure alignment with the desired therapeutic impact.
    \item A RAG should utilize datasets that scientists can trust, such as the STRING dataset.
    \item A RAG should provide the PPI alongside relevant explanations to enable fact-checking. 
\end{itemize}

\begin{table*}[htbp]
\centering
\begin{adjustbox}{width=2\columnwidth,center}
\begin{tabular}{cl}
\hline \hline
\textbf{Notation} & \textbf{Description} \\ \hline
\textit{protein} (Node) & the name of a protein \\  \hline
\textit{start\_annotation} (Node) & the text to describe the property of the start protein in a PPI \\ 
\textit{end\_annotation} (Node) & the text to describe the property of the end protein in a PPI \\ \hline
\textit{embedding} (Node) & the semantic embedding vector of the annotation text \\ \hline
\textit{combined\_score} (Edge) & a score to show the confidence of the PPI \\ 
\textit{interaction\_type} (Edge) & a text to indicate the type of the PPI \\ \hline
\textit{path} (Path) & names of proteins along the PPI signaling pathway  \\ 
\textit{path\_details} (Path) & generated explanations and retrieved annotations of a PPI pathway\\
\hline \hline
\end{tabular}
\end{adjustbox}
\caption{Attributes of the constructed knowledge graph (KG). (Node) indicates the attributes of nodes, (Edge) represents the attributes of edges, and (Path) represents the attributes of paths.}
\label{table:kg}
\end{table*}

\section{Preliminaries}
In this section, we introduce the formulation of the problem and the construction of our medical graph database. We conducted a user study with domain experts in DD to understand their requirements for PPI exploration: a framework that can understand their initial protein and therapeutic impact then outputting recommended PPI pathways with explanations and retrieved contexts from reliable datasets for fact-checking. More details on formative study are provided in the Appendix.


\subsection{Problem Formulation}
Let $\mathcal{G} = (\mathcal{V}, \mathcal{E})$ be an instance of a knowledge graph, where $\mathcal{V}$ is the set of nodes and $\mathcal{E}$ is the set of edges. Node $v \in \mathcal{V}$ and edge $e \in \mathcal{E}$ represent protein and PPI in the PPI network.  The goal of \system is to provide potential PPI signaling pathways and offer AI-generated explanations and database-retrieved contextual information to support experts' decision-making. The following equation can describe the input and output of \system:

\begin{equation}
\begin{aligned}
\text{GraPPI}(G, Q_{\text{users}}, IP) = &\arg\max_{P \subseteq G} \\
\Big( S(P, Q_{\text{users}}, &X_{\text{AI}}(P), I_{\text{DB}}(P)) \Big)
\end{aligned}
\end{equation}

In this equation, $G$ represents the interaction graph derived from the database, while $P$ represents the potential PPI signaling pathway. \textit{IP} represents the initial protein. $Q_{\text{users}}$ indicates the therapeutic impact from users. $I_{\text{DB}}(P)$ represents the retrieved information about certain pathways from the database. The equation maximizes the relevance score $S(P)$ for a pathway based on a given therapeutic impact $Q_{\text{users}}$. $X_{\text{AI}}(P)$ (AI-generated explanation), and $I_{\text{DB}}(P)$ (retrieved information from the database) over the pathway $P$ are utilized as supplementary information for evaluation. We will assess the results of the recommended PPI candidates using quantitative metrics for semantic similarity literal alignment and subjective evaluation of domain experts in DD.



\subsection{Medical Graph Database Construction}
\begin{table}[H]
\centering
\begin{adjustbox}{width=0.9\columnwidth,center}
\begin{tabular}{|l|c|c|}
\hline
\textbf{Dataset} & \textbf{\#Entities} & \textbf{\#Triples} \\ \hline
\textbf{Ours}  & 18,767 & 2,955,220  \\ \hline
\end{tabular}
\end{adjustbox}
\caption{Basic statistics of our knowledge graphs. 
}
\label{table:datasets}
\end{table}

In this section, we describe the process of constructing our KG database containing domain knowledge of all human PPIs from the STRING dataset \cite{szklarczyk2023string}, through data collection and deployment. After domain experts in drug discovery confirmed the reliability of the STRING dataset, we collected information on all proteins within the \textit{Homo sapiens} category, identifying and reorganizing representative features including \textit{combined\_score}, \textit{interaction\_type}, \textit{protein}, \textit{text annotation}, and \textit{embedded-annotation vector}. The \textit{text annotation} and \textit{combined\_score} attributes are sourced from prior work, with their reliability validated by experts.  
Attribute definitions are summarized in Table~\ref{table:kg}. The embedding model used is OpenAI’s \textit{text-embedding-3-small} \footnote{\url{https://platform.openai.com/docs/guides/embeddings}}. We employed a \textit{neo4j} database dump to inject the reconstructed dataset and deployed it on Google Cloud Platform for cloud hosting. The resulting graph database contains 18,767 nodes and 2,955,220 edges, representing the largest PPI graph database for \textit{Homo sapiens} proteins to date, with the most comprehensive PPI relationships.  

\section{\system Framework}


To add supporting information for LLMs to understand single PPI and multi-PPI signaling pathway analysis, we developed a KG-based RAG framework to support PPI pathway exploration. The framework contains two components: (1) \textbf{Interaction Graph: Moving \textit{k}NN windows} to extract the relevant subgraphs from the KG to enable LLM inference over the large-scale KG, and (2) \textbf{Impact Search: A retrieve-divide-solve style agent pipeline} to understand PPI pathways on single PPI edge level and entire PPI pathway level context. The overview of \system is illustrated in Figure \ref{fig:overview}.


\subsection{Interaction Graph: Moving \textit{k}NN Graph Windows} 
After discussion with domain experts, we constructed the entities in KG: <\textit{head protein, combined\_score, tail protein}>. To start to generate \IG as the PPI context, \system will receive the name of the initial protein from users and use Cypher statement to extract all the connected nodes (proteins) in the KG. 
Among all the connected protein candidates, \system will implement a moving \textit{k}-nearest neighbor (\textit{k}NN) graph window strategy to form interaction graphs allowing users to select potential PPIs and mitigate redundant inference in \IG. 
The pseudocode for moving \textit{k}NN graph windows is shown in Algorithm \ref{alg:knn_depth}. 
In this case, we use the FAISS open-source library for fast \textit{k}NN search in high-dimensional dense representation space\cite{johnson2019billion}.

\begin{algorithm}
  \small
\caption{Moving kNN Graph Windows}\label{alg:knn_depth}
\begin{algorithmic}[1]
\Require: Initial protein $p_{\text{init}}$, \textit{k}NN constants $K$, Window indexes $I$, Graph size parameter $M$
\setcounter{ALG@line}{0} 
\Ensure Expanded graph $G$
    \State Initialize interaction graph $G$ with $p_{\text{init}}$ as the root node
    \State Set $m \gets 0$
    \While{$m < M$}
        \For{$protein \in G$}
            \State Determine the moving \textit{k}NN window $[i*k + 1, (i+1)*k], i\in I, k \in K$
            \State Retrieve k connected proteins using the moving \textit{k}NN search of annotation embeddings
            \State Add retrieved proteins and edges to $G$
            \State Increment $m \gets m + 1$
        \EndFor
    \EndWhile
\end{algorithmic}
\end{algorithm}

\subsection{Impact Search: Retrieve-Divide-Solve}
In \IS, users can select the interaction graph that contains their preferred proteins for LLM inference based on therapeutic impact. After interaction graph selection, \IS will take the interaction graph and the therapeutic impact query as the input to perform the PPI pathway exploration using a retrieve-divide-solve style agent pipeline. Instead of directly incorporating long-context annotations in the long PPI signaling pathways, \IS first decomposes the analysis for the entire tasks into multiple subtasks, with each subtask retrieving the long annotations of two end proteins to explain each PPI in parallel for the whole of the PPI pathway recommendation. Next, we will walk through \IS in \system step by step.



\subsubsection{Step I: Edge Explanation}

Instead of directly analyzing the entire PPI pathway, which is very likely to have an extremely long PPI context, the edge explanation agent breaks down the entire path into a collection of edges with detailed information about proteins. Each edge will be analyzed with corresponding protein context and therapeutic impact to generate shorter and more concise explanations with less redundant information. The edge explanations will be the information source for path explanation and supporting context for fact-checking.

\subsubsection{Step II: Path Exploration} 
In Step II, the edge explanation agent
will generate the explanation for the entire PPI signaling pathway.
The generated path explanation and the relevance score will be stored as attributes of each path. 
After the generation, all the intermediate results from \textit{k}NN search in \IG to Edge Explanation and Path exploration in \IS are accessible to domain experts to support their fact-checking.

\subsubsection{Step III: Re-rank} 

When generating path explanations, the path explanation agent is also responsible for evaluating the relevance of the pathway to the therapeutic impact using the demonstrated zero-shot ranking ability of LLMs \cite{hou2024large, zhao2023recommender} since previous studies show the strong potential of transferring knowledge from LLMs as powerful recommendation models. Using the calculated relevance score, \system will show the top \textit{n} relevant pathways among all PPIs to users rather than using the ranking in the previous interaction graph. Users specify the ranking window length \textit{n}.
\section{Experiment}
\begin{table*}[htbp]
\centering
\begin{adjustbox}{width=1.8\columnwidth,center}
\begin{tabular}{l||c||c||c}
\toprule
\multirow{2}{*}{\textbf{Methods}} & \multicolumn{3}{c}{\textbf{metrics (Mean±Std)}} \\
\cmidrule{2-4} 
 & \textbf{BERTScore F1} & \textbf{ROUGE-1 F1} & \textbf{ROUGE-L F1}  \\
\midrule

\textbf{GPT-4o-mini-Baseline} & 86.76±2.13 & 24.30±8.54 & 21.33±6.83  \\
\textbf{GPT-4o-mini-Zero shot w/ CoT} & 88.55±1.99 & 26.38±8.13& 30.96±11.29  \\
\textbf{GPT-4o-mini-RAG w/o CoT} & 87.53±1.61 & 27.80±11.79& 22.51±6.26 \\
\textbf{GPT-4o-mini-Ours} & \underline{88.98±1.32} & \underline{30.41±7.12}& \underline{32.28±6.57}  \\

\midrule
\textbf{GPT-4-Turbo-Baseline} & 88.00±2.00 & 32.46±11.17& 24.40±9.14 \\
\textbf{GPT-4-Turbo-Zero shot w/ CoT} & 89.32±1.58 & 36.69±11.39& 31.02±11.03  \\
\textbf{GPT-4-Turbo-RAG w/o CoT} & 89.95±0.94 & 38.70±7.04 & 31.93±6.61  \\
\textbf{GPT-4-Turbo-Ours} & \underline{\textbf{90.02±0.93}} & \underline{42.19±5.53} & \underline{\textbf{37.47±6.96}}  \\

\midrule
\textbf{GPT-4o-Baseline} & 86.76±2.13 & 29.10±8.28& 21.33±6.84  \\
\textbf{GPT-4o-Zero shot w/ CoT} & 87.64±2.28 & 34.07±12.25& 30.76±10.75 \\
\textbf{GPT-4o-RAG w/o CoT} & 88.70±1.46 & 34.36±4.15& 30.06±4.94  \\
\textbf{GPT-4o-Ours} & \underline{89.62±1.82} & \underline{\textbf{42.26±11.96}}& \underline{35.67±12.58} \\

\bottomrule
\end{tabular}%
\end{adjustbox}
\caption{Performance comparison on different models and configurations. \underline{Underline} means the best configuration in a given model and \textbf{Bold font} represents the top-performing methods in the evaluation. \textbf{Baseline} represents vanilla model setting without retrieved materials from our KG and a co-designed chain-of-thought (CoT) with domain experts. \textbf{Zero shot} means that models will do inference without the support of retrieved contexts from the KG. \textbf{Ours} includes both components to complete the pipeline.}

\label{Table_App}
\end{table*}



We will assess the performance of \system in multiple configurations to answer the following experimental questions (EQs): 
\begin{itemize}
    \item \textbf{EQ1:} To what extent does \system improve the accuracy of path explanations compared with baseline models?
    \item \textbf{EQ2:} How does the retrieve-divide-solve pipeline improve the accuracy and efficiency when scaling up interaction graphs? 
    \item \textbf{EQ3:} How do the domain experts perceive the utility of AI-generated explanations and retrieved information in supporting their decision-making?
\end{itemize}
\textbf{EQ1}, \textbf{EQ2}, and \textbf{EQ3} will be explained in Section \ref{sec:SIE}, \ref{sec:SVE}, and \ref{sec:CS}, respectively. 
\subsection{Setup}
For the evaluation, we accessed the performance of our framework in terms of two aspects: (1 )accuracy to demonstrate how the generated contents align with the requirements of users and the background information, and (2) the scalability that evaluates the framework performance when \IG scales up. We conducted the experiments on both edge-level generated explanation and path-level explanation. For the comparative study and the ablation study, the graph size parameter was set to be 2, and the hyperparameters for \textit{k}NN $k_1$ and $k_2$ to explore the nodes in depth 1 and depth 2 were set to be 10 and 2 respectively. \IS will recommend the top 10 PPI signaling pathway candidates for each interaction graph according to the relevance to the therapeutic impacts from users. For scalability analysis, we test the performance of \system using 6 \IG with different sizes. The depths of those \IG are \textit{2} and \textit{3} while \textit{k} value for \textit{k}NN are \textit{10}, \textit{15}, and \textit{20} respectively. The numbers of paths are set as follows: \textbf{$[40, 75, 76, 95, 113, 160]$}. The \IG in the experiment group is using \system to generate path explanations based on the edge explanations. The \IG in the control group directly refers to raw annotation texts of each protein in the pathway to generate path explanations. 

\subsection{Baseline}
We assess the performance of our framework using three state-of-the-art LLMs as base models: GPT-4o\footnote{\url{https://openai.com/index/hello-gpt-4o/}}, GPT-4-Turbo\cite{achiam2023gpt}, and GPT-4o-mini\footnote{\url{https://platform.openai.com/docs/models\#gpt-4o-mini/}}. GPT-4o and GPT-4o-mini are the cost-effective variants of GPT-4-Turbo designed to balance performance and computational expense.



\subsection{Tasks and Metrics}
\begin{figure*}[h]
    \centering
    \includegraphics[width=\linewidth]{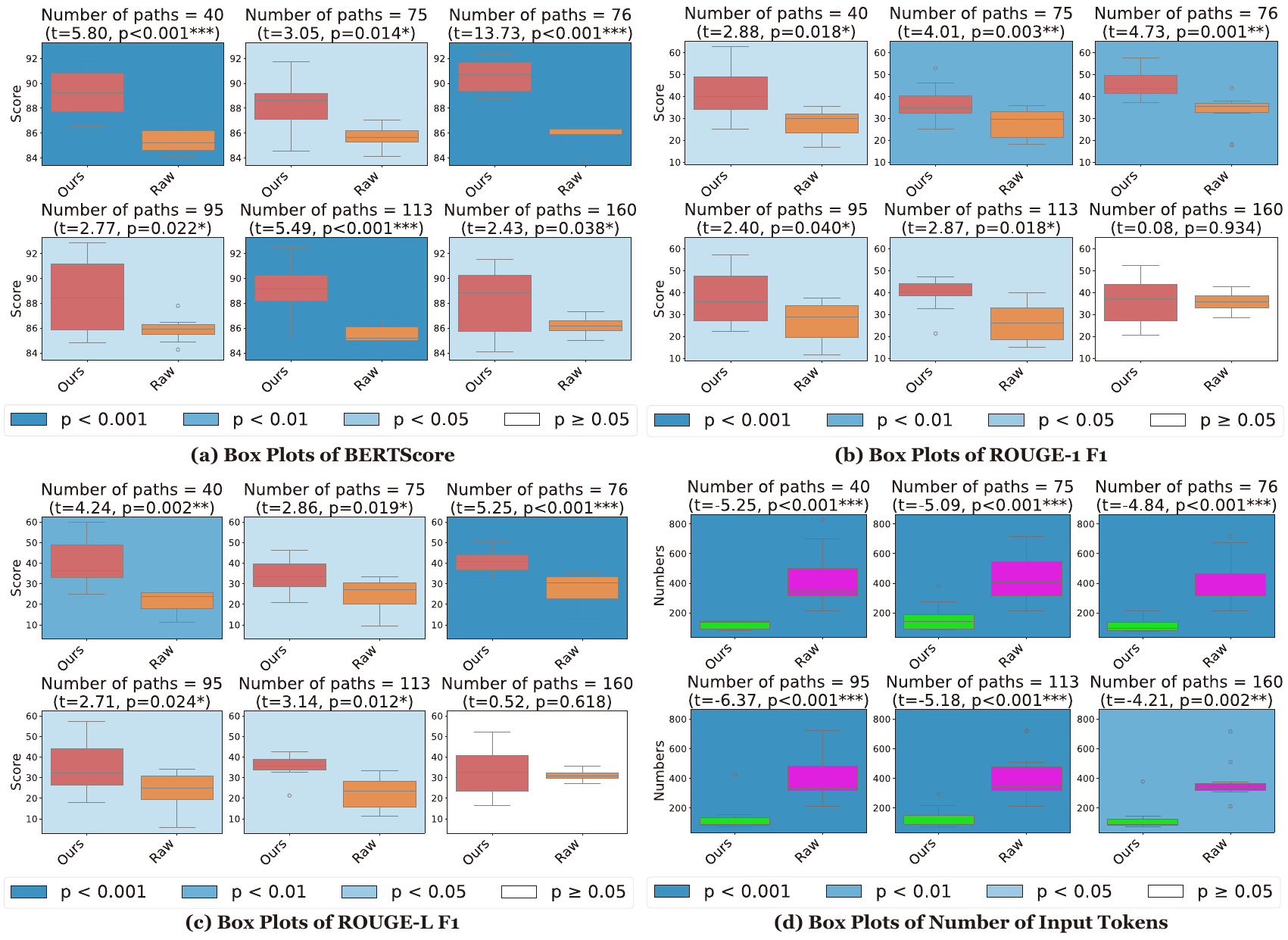}
    \caption{Box plots of results under different graph sizes. The blue color in the background indicates the level of difference between the two groups. Darker blue represents a more significant difference. For plots (a), (b), and (c), the red and orange boxes represent different accuracies of \system and the system directly using protein annotations, respectively. For plot (d), the green and purple boxes indicate the number of input tokens they have. \textbf{Raw} refers to the methods using raw annotations texts as contexts while \textbf{Ours} utilizes the edge explanations with more concise representation of biomedical context. 
}
    
    \label{fig:SVE}
\end{figure*}

To evaluate the accuracy of the generated contents, we employed \textbf{BERTScore} \cite{zhang2019bertscore}, \textbf{ROUGE-1}, and \textbf{ROUGE-L} \cite{lin2004rouge} to assess semantic alignment and literal overlapping. Since our framework aligns the generated content with both retrieved knowledge from our database and users' intent, we combined input query with retrieved information and formed the input tuple of edge level as: <\textit{"query", "start\_protein", "end\_protein", "start\_annotation", "end\_annotation"}>. The input tuple of path level is: <\textit{"query", "path", "path\_details"}>. The definitions of those variables are shown in Table \ref{table:kg}. The metrics are calculated between the input tuples of edges level along the recommended PPI pathway and the generated path explanation. Regarding scalability evaluation, we conducted scale variant evaluation to compare the performance of \system under different scales of \IG shown in Figure \ref{fig:SVE}.

\subsection{Experimental Results}

\subsubsection{Ablation Study}\label{sec:SIE}
\textbf{EQ1: Accuracy}: In this experiment, we conducted the accuracy evaluation of \system under a given data scale. As shown in Table \ref{Table_App}, \system outperforms not only the baseline but all other configurations with all three base models in terms of semantic similarity (BERTScore) and literal alignment (ROUGE-1 and ROUGE-L). Four configurations are designed: Ours, RAG w/o Chain-of-Thought (CoT), Zero shot w/ CoT, and Zero shot w/o CoT (Baseline). To ensure consistency, we implemented those configurations on the same initial protein from domain experts for sub-graph retrieval, and we adjusted LLM inference settings for the ablation study. One interesting thing we observed while manually checking the generated explanations is that although GPT-4-Turbo has the best quantitative performance among all three base models, its explanations may prioritize protein annotation details over user-specified therapeutic impacts. GPT-4-Turbo generates explanations covering more divergent topics, while GPT-4o focuses more on the users' query to give more concise explanations.  

\subsubsection{Scale Variant Evaluation}\label{sec:SVE}

\textbf{EQ2: Scalability and Efficiency}: We evaluate the performance of \system using GPT-4o under interaction graphs of different scales. As shown in Figure \ref{fig:SVE}, by comparing all four metrics in the six scales, significant differences exist between the two data groups for all metrics except ROUGE-1 and ROUGE-L when the number of pathways reaches 160.
In this experiment, six \IG containing different numbers of pathways are used to test \system's scalability. For each \IG, \system recommends 10 potential PPI pathway candidates based on their relevance to user queries. We employed paired t-tests to determine whether \system's introduction leads to significant performance differences. While no statistically significant differences were observed between groups for ROUGE-1 and ROUGE-L at 160 paths, the mean scores for \system still surpassed those of the Raw group, indicating superior literal alignment. Overall, \system achieves better semantic similarity and lexical alignment while using significantly fewer tokens. These results highlight \system's potential as an accurate and token-efficient computational tool for exploring PPI pathways in large-scale datasets.


\subsubsection{Case Study}\label{sec:CS}
\textbf{EQ3: Explainability}: We further conducted a case study to evaluate the quality of responses generated by \system. Users reviewed the generated content in \IG and \IS to assess \system's contribution to exploring novel PPI pathways. We interviewed users about their current \textit{Target ID} practices and the applicability of LLMs in \textit{Target ID} to validate the generated content. Given an initial protein and therapeutic impact, \system produced 5 PPI pathways (containing 9 PPI edges) with explanations and retrieved descriptive texts for pathway validation. Users leveraged these outputs to explore new PPI pathways. All generated explanations were deemed coherent, and users could integrate them with prior work context to make decisions. Baseline model outputs were less convincing due to the lack of fact-checking support, with some requiring manual calibration. \system offers a novel approach to PPI pathway exploration, providing biomedical explanations and retrieved information to mitigate over-reliance on AI. Case study details and sample results are provided in Figure~\ref{fig:case} in the Appendix.  
\section{Related Works}


\subsection{LLM-based and RAG-based systems in PPI prediction}
Large Language Models (LLMs) have demonstrated significant potential in understanding and generating natural language responses \cite{wang2024biorag, li2023gpt4rec, kang2023llms}. Inspired by LLMs, pioneering work has focused on building protein language models (PLMs), which are pre-trained on large-scale protein sequences \cite{hsu2022learning, elnaggar2021prottrans, li2023understand}. PLMs capture more accurate protein features by representing sequences as high-dimensional embedding vectors. Previous studies have leveraged PLMs to enhance performance in downstream tasks such as protein structure prediction \cite{lin2023evolutionary} and PPI prediction \cite{jin2024prollm}. However, while PLMs exhibit strong capabilities in biomedical tasks, they remain prone to hallucinations—generating text that is nonsensical or unfaithful to source content \cite{ji2023survey, zhang2023siren}. To address this, Retrieval-Augmented Generation (RAG) has been introduced to mitigate hallucinations and improve trustworthiness by retrieving contextual information from external databases \cite{khandelwal2019generalization, lewis2020retrieval}. RAG enhances the reliability and transparency of biomedical LLM-based systems \cite{yang2024harnessing}.  


\subsection{Explainable AI} 

Heated discussions have centered on explainable AI (XAI), which generates actionable explanations for AI outputs instead of treating LLM-based systems as black boxes \cite{wiegreffe2021teach, schuff2022human, lamm2021qed}. Transparent LLM-based systems that provide explanations and retrieve materials from existing research enhance human verification of AI-generated content \cite{vasconcelos2023explanations}, assist in decision-making \cite{fok2023search}, calibrate user trust \cite{bussone2015role}, and reduce over-reliance on AI \cite{das2022prototex, zhang2020effect, bansal2021does}. Thus, providing reliable background information is critical to mitigating over-reliance on AI and enabling experts to calibrate trust through explanations and retrieved context \cite{si2023large}.  

\section{Conclusion}
In this paper, through a carefully designed user study with drug discovery researchers, we observed the importance of retrieving PPIs with the therapeutic impact they target in \textit{Target ID} (i.e., find PPIs to inhibit MARK4). Therefore, we proposed \system, which is a large-scale KG-based retrieve-divide-solve style agent pipeline RAG framework to support PPI pathways exploration in understanding therapeutic impacts with two key components: 1) moving \textit{k}NN windows for sub-graph extraction, and 2) a retrieve-divide-solve style agent pipeline to incorporate long protein annotations for inference.
Through extensive experiments, we demonstrate that: (1) \textit{accuracy}: \system consistently outperforms baseline methods in semantic and lexical alignment; (2) \textit{scalability and efficiency}: \system achieves comparable results using significantly fewer tokens than raw annotation-based systems, and (3) \textit{explainability}: \system generates more reliable explanations and retrieved information, validated by domain experts.  
\section{Limitations}

In this paper, we developed a KG-based RAG framework to support PPI signaling pathway exploration. However, our study has several limitations. 

\textbf{First}, while domain experts in DD expressed full confidence in the STRING dataset during interviews, the dataset does not comprehensively cover all known PPIs, as biomedical research continues to discover new interactions, and the STRING dataset requires constant updates. During the case study, one expert noted that protein \textit{TUBB3} interacts with \textit{MAPT} via phosphorylation based on their expertise, but this interaction is absent in STRING. Integrating features such as PPI prediction models or allowing researchers to input custom findings could enhance dataset coverage. 

\textbf{Second}, the case study utilized only two initial proteins validated by DD experts. While these proteins were well-understood, they do not represent the full diversity of protein interactions, leaving potential edge cases unexamined. Future work should incorporate a broader range of initial proteins to rigorously assess \system's applicability.

\textbf{Third}, while the retrieve-solve-merge pipeline improves PPI exploration at moderate scales, its benefits diminish as interaction graphs grow larger. This is due to the increasing length of edge explanations approaching LLM context window limits. Thus, \system is not a universal solution for arbitrarily large graphs; users must judiciously set the \textit{k}NN window size to balance exploration scope and explanation quality. This approach not only generates smaller interaction graphs for analysis but also prioritizes shorter PPI pathways, which are clinically preferable due to reduced side effects and more efficient signaling transmission.

\bibliography{acl_latex}

\appendix

\section{Appendix}\label{sec:appendix}
\begin{figure*}[htbp]
    \centering
    \includegraphics[width=0.999\linewidth]{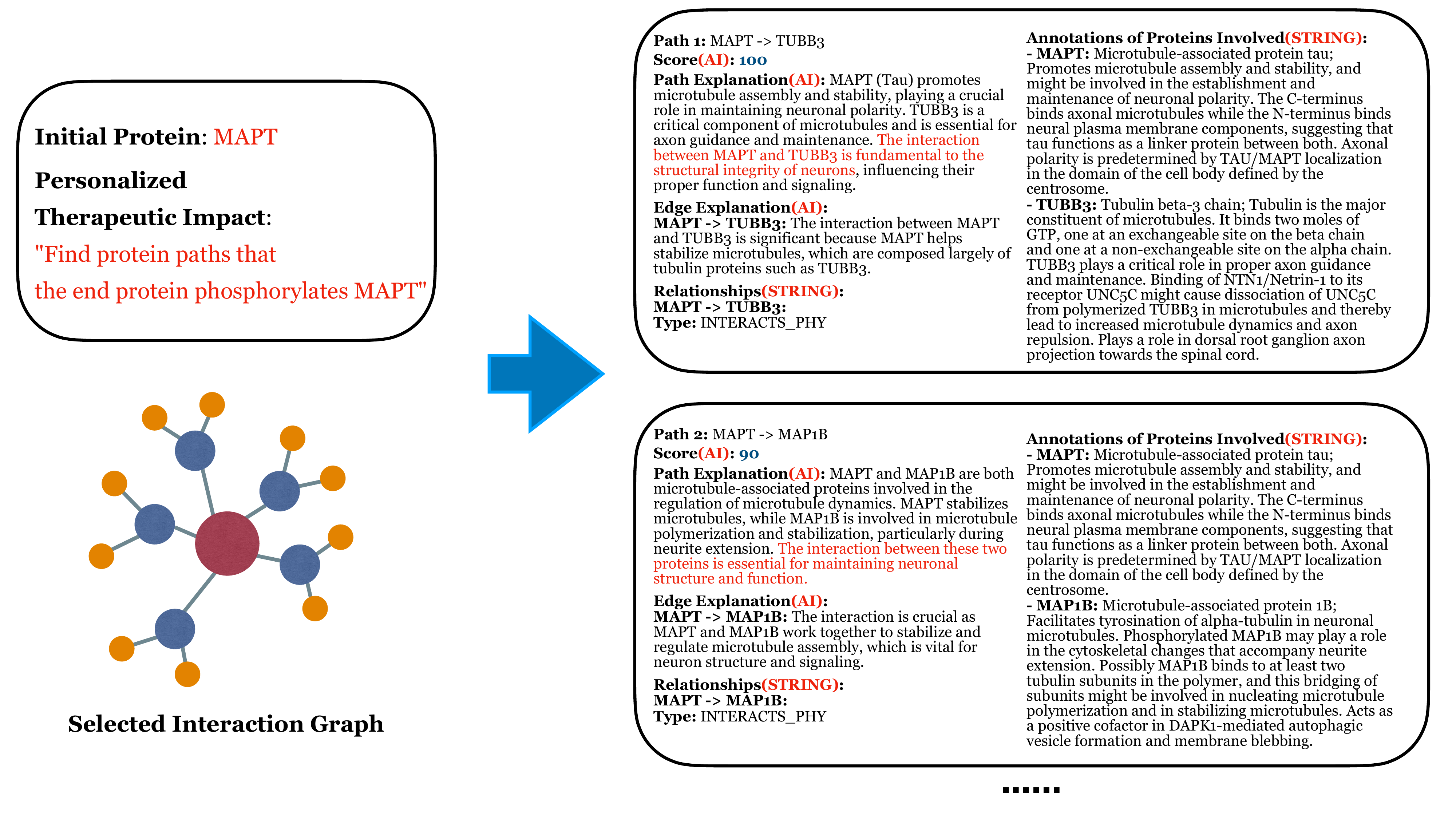}
    \caption{Part of the results of the case study showing the input and output content regarding certain recommended PPI signaling pathways. }
    \label{fig:case}
\end{figure*}

\begin{table*}[ht]
\centering
\caption{Demographic distribution of the five user study participants. The table includes details about gender, age range, profession, and work experience.}
\begin{tabular}{>{\raggedleft\arraybackslash}m{2cm}>{\raggedleft\arraybackslash}m{4cm}||cc}
\toprule
\multicolumn{2}{c}{{\textbf{Demographic group}}}                                & \multicolumn{2}{c}{\textbf{User Study}}\\
\multicolumn{2}{c}{}                                                                           & \textbf{N}     & \textbf{\%}\\
\hline
\midrule
\multirow{2}{*}{Gender}                  & Female       & 1 & 25\%  \\
                                         & Male         & 4 & 75\%  \\
\midrule
\multirow{2}{*}{Age Range}               & 26-35        & 3  & 60\%  \\
                                         & 36-45        & 2  & 40\%  \\
\midrule
\multirow{2}{*}{Profession}              & Post-Doctor      & 2 & 40\%  \\
                                         & Researcher                   & 3 & 60\%  \\
\midrule
\multirow{2}{*}{Work Experience}         & 7-9 years            & 2 & 40\%  \\
                                         & over 10 years        & 3 & 60\%  \\

\bottomrule
\end{tabular}
\label{tab:demo_p}
\end{table*}

\begin{figure*}[htbp]
    \centering
    \includegraphics[width=\linewidth]{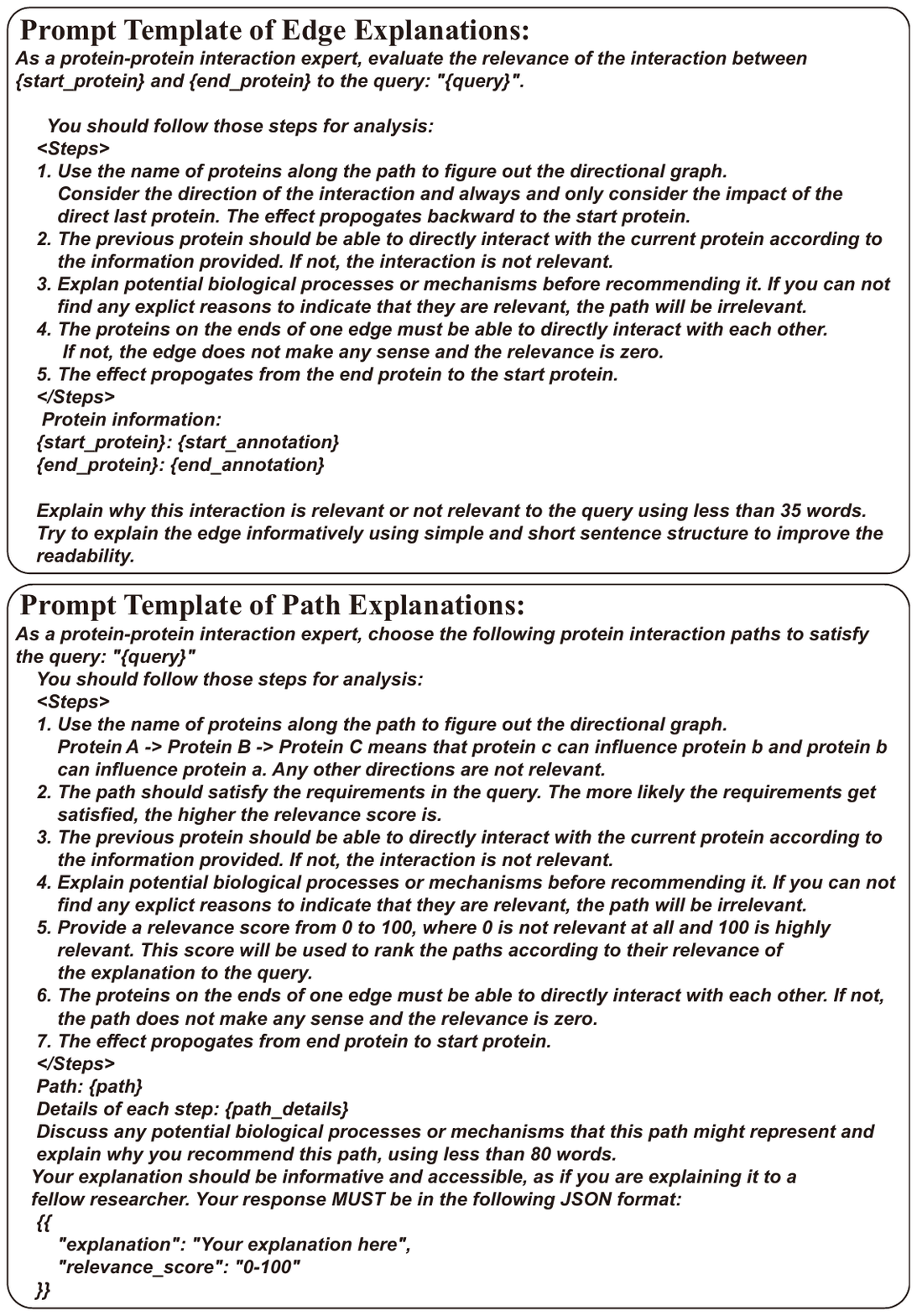}
    \caption{Prompt Templates of Edge Explanation and Path Explanations}
    \label{fig:Prompt}
\end{figure*}

\end{document}